\documentclass[10pt,conference]{IEEEtran}
\usepackage{xcolor}
\usepackage{graphicx}
\usepackage{fancyhdr}

\ifCLASSINFOpdf
\else
\fi
\makeatletter
\def\footnoterule{\kern-3\p@
  \hrule \@width 2in \kern 2.6\p@} 
\makeatother
\newcommand{\copyrightnotice}[1]{{%
  \renewcommand{\thefootnote}{}
  \footnotetext[0]{#1}%
}}

\begin{document}
%
\title{Modeling configuration-performance relation \\in a mobile network: a data-driven approach}

\author{\IEEEauthorblockN{Michał Panek}
\IEEEauthorblockA{Wroclaw University\\of Science and Technology\\
Wroclaw, Poland\\
and Nokia Solutions and Networks\\ Wroclaw, Poland\\
Email: michal.panek@nokia.com}

\and
\IEEEauthorblockN{Ireneusz Jabłoński}
\IEEEauthorblockA{Brandenburg University of Technology,\\Cottbus-Senftenberg, Germany\\
Fraunhofer Institute for Photonic Microsystems,\\Cottbus, Germany\\ Email: ireneusz.jablonski@ipms.fraunhofer.de}
\and
\IEEEauthorblockN{Michał Woźniak}
\IEEEauthorblockA{Wroclaw University\\of Science and Technology\\
Wroclaw, Poland\\Email: michal.wozniak@pwr.edu.pl}}

\maketitle
\copyrightnotice{©2024 IEEE. Personal use of this material is permitted. Permission from IEEE must be obtained for all other uses, in any current or future
media, including reprinting/republishing this material for advertising or promotional purposes, creating new collective works, for resale or
redistribution to servers or lists, or reuse of any copyrighted component of this work in other works.}

\begin{abstract}

Mobile network performance modeling typically assumes either a fixed cell's configuration or only considers a limited number of parameters. This prohibits the exploration of multidimensional, diverse configuration space for, e.g., optimization purposes.

This paper presents a method for performance predictions based on a network cell's configuration and network conditions, which utilizes neural network architecture. We evaluate the idea by extensive experiments, with data from more than 50,000 5G cells. The assessment included a comparison of the proposed method against models developed for fixed configuration. Results show that combined configuration-performance modeling outperforms single-configuration models and allows for performance prediction of unknown configurations, i.e., it is not used for model training. A substantially lower mean absolute error was achieved (0.25 vs. 0.45 for fixed-configuration MLP-based models).

\end{abstract}

%
\IEEEpeerreviewmaketitle

\section{Introduction}
With the growing mobile network's complexity, measured in terms of the number of network elements and the size of configuration space \cite{Imran2014}, network management inevitably moves towards a greater level of automation. Autonomous management of network performance is currently realized mostly by detecting anomalous behaviors that greatly stands out from past values or deviates from current trends \cite{Moysen2020}. Rarely it is considered in conjunction with the network element's configuration or network conditions. Recently, the topic of Network Digital Twin (NDT) has been gaining attention \cite{Vila2023}. NDTs are proposed to be used for network optimization, where configuration changes are first assessed in virtual space before being applied to physical network elements. It is deemed to bring advantages by, e.g., allowing fast and safe experiments without a thread of negative impact to end users \cite{almasan_network_2022}. Yet, to our best knowledge, no works have addressed the issue of combining broad configuration space and modeling its impact on performance. This topic is vital for NDT or automation of performance management. Currently, if considered at all, only a limited number of factors are considered. To address this gap, we present our works on modeling the relationship between cell Key Performance Indicators (KPIs) with both environmental factors experienced by a given cell and its configuration. To the best of our knowledge, this is the first work in which the relation between a cell's configuration and performance is modeled with a single data-driven model capable of making predictions for multiple configurations described by more than 100 parameters.  

The contribution of this paper is as follows: We propose the configuration-performance (CMPM) modeling approach and describe the respective pipeline for mobile network data processing. The proposed approach has been validated with experiments utilizing live 5G network data.

\section{Background and related works}

\subsection{KPI modeling}

Early approaches of performance modeling were based on auto-regressive methods, e.g. Auto-regressive Integrated Moving Average (ARIMA) or Fractional ARIMA (FARIMA) models \cite{Sang_ARMA}. For forecasting purposes, especially with low-sampling rate and aggregated data, these methods provide satisfactory results. Recent years are marked with proliferation of machine learning in science and engineering, which has been used for the task of performance prediction using algorithms like AdaBoost, Support Vector Regression (SVR) or Neural Networks (NN) \cite{Moysen2020,Parera2021}. Long Short Term Memory (LSTM) have gained a lot of popularity \cite{Kurri2021}. However, all those approaches were auto-regressive ones, where prediction was based on lagged values of the same Key Performance Indicator. Other research works proposed to leverage multiple features to model network performance. Sliwa and Wietfeld \cite{sliwa2019} models link throughput using contextual information on instantaneous radio conditions, terminal velocity and data volume to be transmitted. In \cite{Elsherbiny2020}, radio transmission quality measures like \textit{reference signal received power} (RSRP) and \textit{reference signal received quality} (RSRQ), enriched with spatio-temporal data (timestamp and location data), were used to predict instantaneous throughput. The relation between throughput KPI and 'environmental' metrics has also been  considered in a number of works (\cite{roy_quantifying_2016,Nikravesh2016,Mostafa2022,Mendoza2023}), where multiple metrics have been used to predict throughout, e.g. \textit{block error rate(BLER)}, \textit{channel quality indicator (CQI)}, \textit{data volume}, \textit{physical resource block (PRB)  utilization} and \textit{number of active user equipments (UEs)}.

\subsection{Configuration-Performance Relation}
In software systems, it is a common practice to predict performance from configuration data \cite{Shailesh2018, Ha2019}. 
Commonly, such methods do not take into account elements beyond configuration parameters, primarily because performance evaluations are conducted in consistent, controlled settings. In mobile networks, this approach was considered e.g. in \cite{Shodamola2020}, where authors are using two parameters (\textit{cell individual offset} and \textit{handover margin}) to predict SINR in handover situation. A combination of configuration parameters and environmental factors has also been considered in literature. Authors of \cite{Adeel2015} use predicted per-UE throughput values to optimize overall performance of LTE base station. They propose to use Random Neural Network, fed with a mix of environmental data (\textit{SINR} and \textit{Inter-Cell Interference}) and per-scheduled transmission configuration parameters (\textit{Transmit Power}, \textit{Modulation and Coding Scheme}). 
 
Bhorkar et al. in \cite{bhorkar2019deepauto} proposes to predict KPIs with architecture based on LSTMs and Feed-Forward Neural Network, but enriches the inputs with cell configuration and external data. However, they mention only several parameters: \textit{Cell Band}, \textit{Bandwidth} and \textit{Transmit Power}. 

Authors of \cite{aurora2022} combines configuration and performance data to obtain best performing configuration for given base station. They do not explicitly model performance metrics based on configuration data, but recognize this topic as important to consider in future research.

Nevertheless, the use of configuration data in performance modeling is restricted to only a few factors, thereby reducing the versatility of the approaches introduced.

\section{KPI Modeling - principles}

Let us assume that the relationship between environmental metrics and network KPIs remains constant for a fixed configuration (defined as a combination of configuration parameter values). Thus, the goal is to propose a model that can predict the expected KPI value for a given configuration and given network conditions (environmental metrics). Such modeling is useful for performance management tasks, such as monitoring of KPI deviation after reconfiguration or for NDTs in optimization processes. It allows to examine how given cell would perform in real conditions, but with alternative configuration. In this paper we are exclusively focusing on Downlink Throughput KPI. 

\begin{figure}
\includegraphics[width=3.4in]{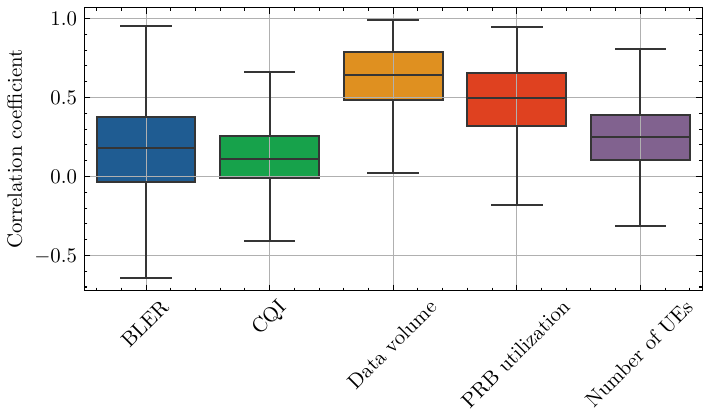}
\centering
\caption{Distributions of Spearman's correlation coefficients among environmental metrics and downlink throughput.}
\label{fig:DL_env_vs_Tput_corr_box}
\end{figure}

Numerous studies have explored the connection between environmental factors and throughput \cite{roy_quantifying_2016,Nikravesh2016,zhang2018lte,sliwa2019,Mendoza2023,minovski2023}. Considering these references and the limitations of the data we could gather, we decided to concentrate on the following environmental indicators: \textit{block error rate (BLER)}, \textit{channel quality indicator (CQI)}, \textit{data volume}, \textit{physical resource block (PRB) utilization}, and \textit{number of active user equipment (UEs)}. The distribution of correlation coefficients for each cell-state in the dataset is shown in Fig. \ref{fig:DL_env_vs_Tput_corr_box}. The highest correlations were observed for \textit{data volume}, \textit{PRB utilization}, and \textit{number of UEs}.

Using all of the configuration parameters would unnecessarily complicate the model as the amount of parameters per cell reaches thousands. What is more, only a subset of parameters are impacting KPIs. For example, operability or security-related variables, identifiers or addresses of neighboring network elements are not directly contributing to achievable performance. Thus, a limited set of parameters can be used both for tracing configuration changes as well as in the modeling tasks. Drawing on the insights from \cite{bhorkar2019deepauto, aurora2022}, a selection of cell parameters was made. Next, we engaged with network configuration specialists to enrich this list with crucial cell-level parameters that could affect the targeted KPI (downlink throughput). Consequently, the following groups of parameters were incorporated into the configuration data subset:
\begin{itemize}
    \item Basic cell's configuration parameters, which includes \textit{bandwidth, band, duplex type, beamforming state (on/off), frame Structure, fronthaul interface type, radio module type, baseband module type, software release, deployment type (NSA/SA)}.
    \item Status of network functionalities, indicating whether specific cellular network features are activated or deactivated. With every new release of network element's software, equipment vendors are providing new enhancements (denoted here as functionalities) which are aimed at providing new capabilities to the network. Examples are: \textit{256QAM modulation in downlink} or \textit{scheduler enhancements}. Functionalities were selected based on categorization provided by the equipment vendor, which divides set of all functionalities into categories based on expected effect. 
\end{itemize}

\begin{figure}
\includegraphics[width=3.4in]{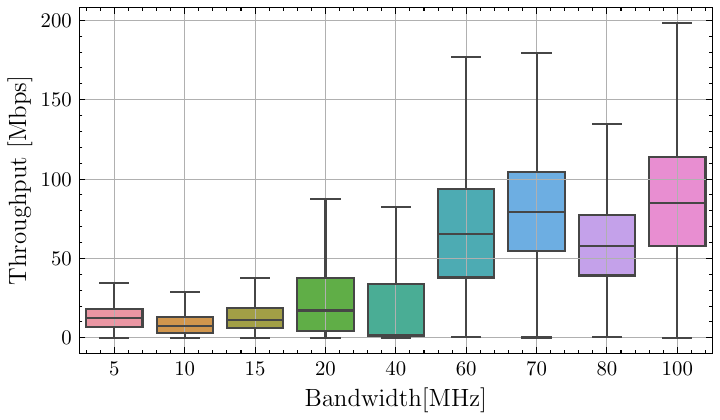}
\centering
\caption{Distributions of downlink throughput per band recorded in live 5G network.}
\label{fig:DL_Tput_box_per_band}
\end{figure}

For example cell's bandwidth defines the number of resources available for data transmission. The higher the bandwidth, the higher the throughput can be. This is illustrated on Fig.\ref{fig:DL_Tput_box_per_band}, where distributions of downlink throughput aggregated over cells with the same bandwidth are drawn.

\subsection{CMPM modeling}
We are utilizing data collected from more than 50,000 5G cells with actual traffic, located in various geographical areas, over a period of 5 months. This data is captured from the Operations Support System (OSS), a central entity responsible for managing the network configuration and performance. The nature of the data influences the design of the processing pipeline, and the key attributes are detailed in the following two subsections.
\subsubsection{Configuration Data}

Configuration data is sourced from the Configuration Management (CM) subsystem, which collects configuration details of network elements. The data for an individual Base Transceiver Station (BTS) is typically organized in a hierarchical tree structure, with nodes symbolizing logical units (such as cells within the base station or hardware components) and leaves indicating the specific parameters and their settings. This CM data provides a comprehensive depiction of the configuration of network elements. However, it only reflects the configuration at the moment the snapshot is taken and lacks historical data on parameter changes. To monitor changes in parameters, it is necessary to compare successive snapshots. CM data dumps are produced on a regular basis, but the frequency is determined by balancing business requirements with resource usage. This frequency could range from hours (for a small number of network elements) to weeks (for entire networks in less critical scenarios). A single base station may have up to several thousand parameters.

\subsubsection{Performance Management data}
Data related to performance is gathered from the Performance Management (PM) subsystem. This data includes counters and key performance indicators (KPIs). Counters record the occurrence of specific events within network elements, such as the number of successful handovers or the number of signal quality assessments. These counters serve as the basis for creating KPIs, which provide qualitative insights into the performance of the network. For example, KPIs might measure the success rate of handovers or the average recevied signal's quality. Both counters and KPIs are structured as time series data, typically with a granularity between 15 and 60 minutes, and are aggregated at either cell or BTS level.

We are following a division between environmental metrics and KPIs. From the PM subsystem point of view, these are the same kind of data. We only introduce this division to emphasize which performance metrics provides a description of network conditions (environmental metrics), and which are service performance metrics (KPIs).

PM data are required to keep the same sampling rate across the whole dataset.

\subsection{Data Processing}

\begin{figure}
\includegraphics[width=3.4in]{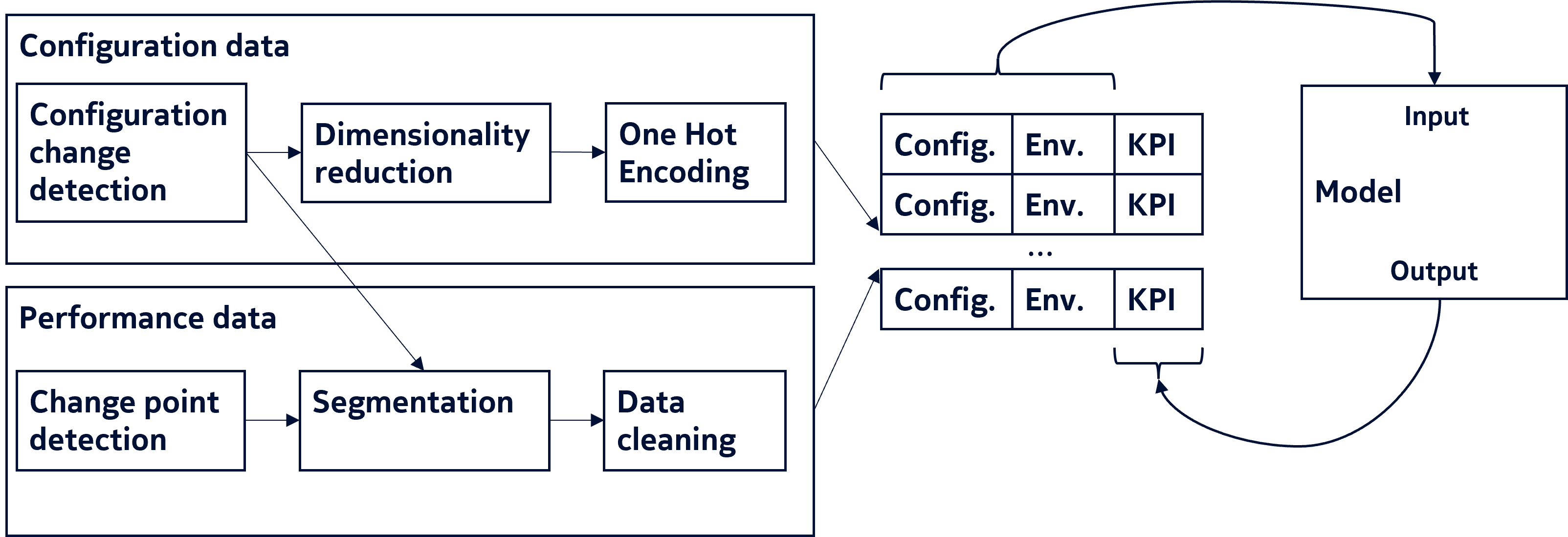}
\centering
\caption{Data processing pipeline for KPI modeling.}
\label{fig:CMPM_processing}
\end{figure}

The data processing pipeline, illustrated in Fig. \ref{fig:CMPM_processing}, is essential for converting OSS data into a usable data set and to develop a KPI prediction model. The initial step involves detecting configuration changes by comparing successive configuration samples of each cell. The timestamps from these changes are utilized to establish stable periods, which are presumed to extend from the detected change timestamp to the timestamp of the final consecutive snapshot with the same configuration. Subsequently, the dimensionality of CM data is minimized by eliminating unnecessary features. This step is executed solely on a subset of CM data that reflects the logical states of functionalities and involves the exclusion of features with only single value in a dataset and features highly correlated (from group of highly correlated features only one is left). The rationale for removing redundant features is that they do not provide additional information to the model, yet they contribute to increased complexity. This can lead to a decline in model performance or increased computational demands. 
Next, a standard one-hot encoding procedure is performed for categorical data.

Processing of performance data starts with an optional step of\textit{ change point detection}. Its aim is to detect abrupt changes in PM time series. When the CM data sampling frequency is low (i.e. the time period between consecutive samples is long), change point detection in PM data may be used to estimate the exact time of configuration change. 
We adopt a safe approach, where the change point detector is set to trigger even for small changes. We have used well-known PELT \cite{Killick2012} for the detection task. We motivate this operation by the observation, that the configuration change either caused an abrupt change in KPI behavior or remained unnoticed.  

\begin{figure}
\includegraphics[width=3.4in]{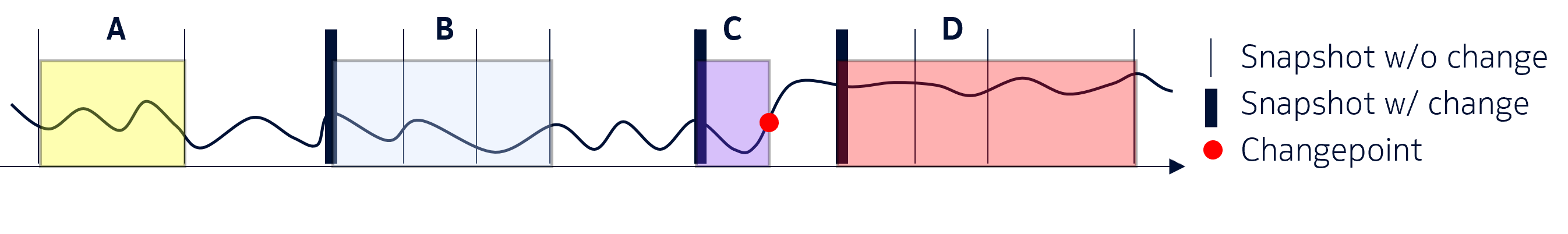}
\centering
\caption{Data segmentation with configuration data and change point detection in performance data. Shaded areas denote periods, where configuration is assumed to be stable.}
\label{fig:segmentation}
\end{figure}

The \textit{Data Segmentation} block is responsible for dividing PM data into chunks with a fixed configuration state. The input from \textit{configuration change detection} is utilized by segmenting PM data by stable periods. Optionally, 
\textit{change point detection} results may be used. Detected change point can expand the stable period. If no change point is detected, we filter out the data for unstable period. Such an operation allows the dataset available for model training to be enlarged, especially in a situation of multiple subsequent changes of configuration. Without change point detection, data in intervals between subsequent configuration changes cannot be used. 
The example of the \textit{Data Segmentation} block results with the use of detected change points is depicted in Fig.~\ref{fig:segmentation}. Stable periods (A, B, D) have been detected by comparing CM data dumps. Interval C lies between two snapshots with configuration changes, however due to change point detected, part of the inter-snapshot data can be used for modeling.

\textit{Data cleaning} operates on data after segmentation. Intervals, for which configuration has not been assigned, are removed. Next, CM data undergoes dimensionality reduction by pruning of highly correlated features, such that only single feature is left from identified group of correlated ones. On the PM side, any missing data are pruned such that whole row is removed. 

\textit{Data merging} is a final step before dataset is created. PM and CM data are joined such that for every row of PM data (consisting of environmental metrics and KPI recorded in a given timestamp for given cell) a respective CM data are appended. Thus, each row of resulting CMPM dataset consist of 3 parts: environmental data, configuration parameters and KPI value. 

For the sake of experiments, we have created a dataset consisting of almost 25 million records with data from over 50000 cells with more than 2400 unique configurations. 12 basic cell's parameters and 114 functionalities were used at input, along with 5 environmental metrics. 

\section{Results}
\textit{Downlink cell throughput} was predicted during experiments. Below the experimental protocol is described, followed by presentation of results. 

\subsection{Experimental Protocol}

\subsubsection{Train-test split}

To correctly assess prediction quality, the relevant experiments shall be thoroughly planned and the split to train/test datasets shall be done with caution. We are considering 2 test cases: prediction for cells with configuration not present in training set and prediction for configurations present in training set, but for cells not included in training set (in any configuration state). Test dataset included almost 900 configurations not present in training set and over 700 overlapping configurations (in this case, the 5G data was collected in various cells across the network). 
 
\subsubsection{Baseline}

\begin{figure}
\includegraphics[width=3.4in]{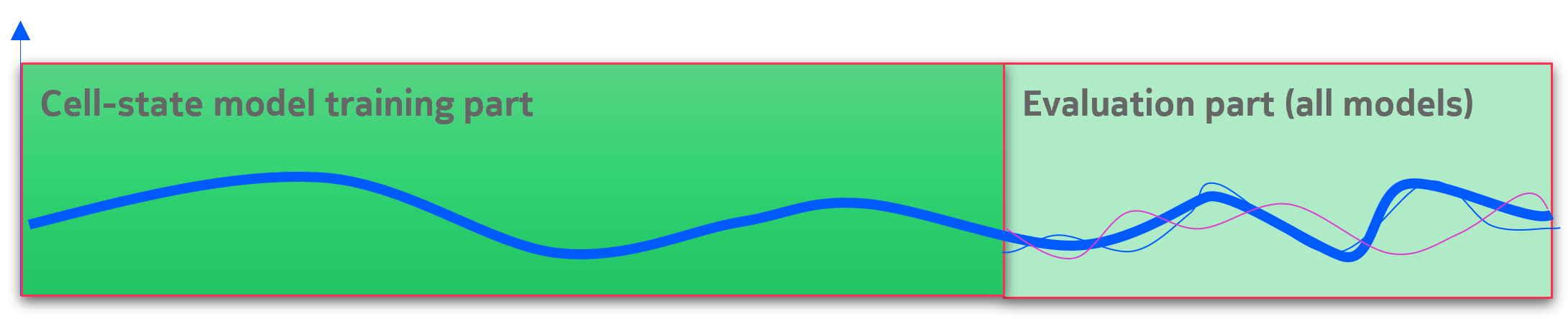}
\centering
\caption{Illustration of evaluation approach. Whole block incorporates data from single cell-state. First 70\% of data is used to train cell-state models (with only environmental data). Remaining 30\% is used to evaluate both cell-state and CMPM models.}
\label{fig:evaluation_NN_models}
\end{figure}

Primary research question is whether prediction with CMPM model is better than those of per-cell models. For each cell-state (i.e. given cell in given configuration state) we have trained per-cell models with only environmental data from this particular cell (no configuration data at model input). Configuration data can be skipped, as within one cell-state configuration remains constant. We have used 70\% of data points from each cell state for that. Four algorithms were used: Ordinary Least Squares (OLS), Multi-layer Perceptron (MLP), AdaBoost and SVR. In the MLP model, a grid search approach was utilized to identify the optimal hyperparameters. The architecture had either one or two hidden layers; in the case of one layer, the allowed neuron quantity was 30, 50, or 100 neurons, whereas in the two-layer scenario, each layer had 30 neurons. The learning rate was fixed at 0.001 or adaptively decreased. The activation function used was ReLu. For the AdaBoost model, a Decision Tree Regressor was employed as the base model with 50 estimators and a learning rate was equal to 1. In the SVR model, an rbf kernel was applied, and the gamma coefficient was determined by the inverse of the product of the training set's variance and the number of features. Evaluation for all modeling approaches was performed on remaining 30\% of each cell-state's data. An example of testing set division for single cell-state is presented in Fig. \ref{fig:evaluation_NN_models}. Note that CMPM model was evaluated with the same approach, i.e. only on the remaining 30\% of data for each cell-state.

\subsubsection{Evaluation metric}
Since the evaluation is done on per cell-state basis, and there are substantial differences in terms of recorded throughput values between cells, the evaluation metric has to be selected such that results can be compared among different cells-states. We have selected Mean Absolute Error (MAE), but calculated it on standardized throughput values and predictions. Standardization was performed for each cell-state separately. 

\subsection{CMPM Model}

We are utilizing fully connected neural network architecture, consisting of 3  layers with 128, 128 and 32 neurons respectively, and Rectified Linear Unit (ReLU) activation function. We used such a simple architecture, as the goal was not to develop the best solution, but rather to assess whether such CMPM modeling is able to outperform single-configuration models. Hereafter, we denote this as CMPM model.

\subsection{Experiments results}
\begin{figure}
\includegraphics[width=3.4in]{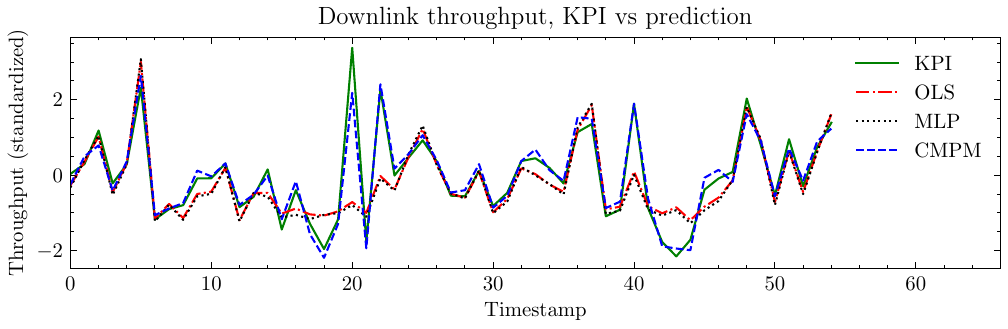}
\centering
\caption{Example of recorded throughput vs predictions for cell-state (OLS, MLP) and CMPM model. KPI values are standardized.}
\label{fig:DL_Tput_VS_Pred_2}
\end{figure}

An example of predictions for one exemplary case is presented in Fig. \ref{fig:DL_Tput_VS_Pred_2}. KPI values are denoted with green solid line, while predictions are depicted with dashed or dotted line. Below, we present analysis of summarized results.

\subsubsection{Known vs Unknown configurations}

First experiments were aimed at evaluating the model's ability to predict throughput for unknown configurations (i.e. configurations not present in training set). The results depicting mean absolute error distributions for known and unknown configurations are presented on Fig. \ref{fig:DL_CDF_MAE_NN_vs_ols_known_unknown_configs}. There is only a tiny difference noticed between cumulative distributions, which leads to a conclusion, that model's prediction quality is similar for both known and unknown configurations.

\subsubsection{CMPM model vs cell-state models}

\begin{figure}
\includegraphics[width=3.4in]{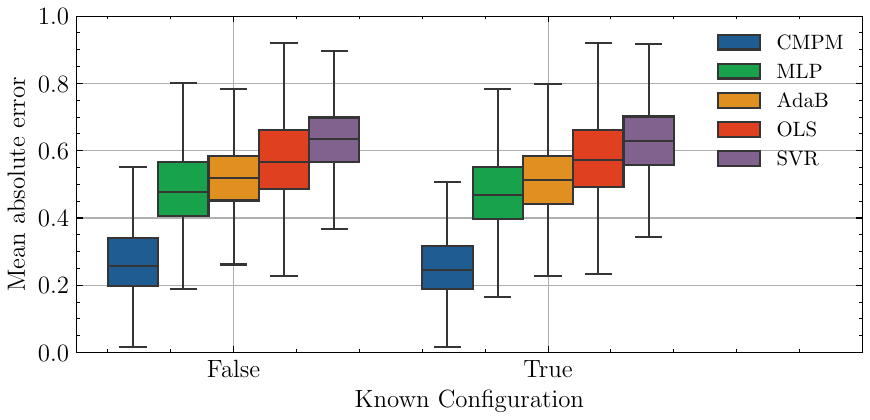}
\centering
\caption{MAE NN vs local models}
\label{fig:DL_boxplots_MAE_NN_vs_ols_known_unknown_configs}
\end{figure}

This experiment was designed to evaluate the effectiveness of CMPM modelling with single model as opposed to models created for each cell state. Since those models do not require implicit knowledge on configuration, the input space is limited from over 100 to 5 dimensions. CMPM models are only useful if able to predict for unknown configurations and at the same time if prediction quality is better than for cell-state models. Results, in form of boxplots for different modeling approaches, are presented on Fig. \ref{fig:DL_boxplots_MAE_NN_vs_ols_known_unknown_configs}. MLP-based models provides lower MAE than OLS, SVR and AdaBoost, but all are outperformed by CMPM model by a great margin. 

The benefit comes from a larger dataset, made possible by the inclusion of configuration data. Cell-state models are limited to training on a small dataset because they are confined to data from a single cell in a specific configuration state for only a brief time. In contrast, the CMPM model capitalizes on the opportunity to use data from various cells with different configurations. Furthermore, incorporating setup parameters enables the model to understand their effects on performance, facilitating predictions for combinations not seen during training.

\subsubsection{Results stability}

\begin{figure}
\includegraphics[width=3.4in]{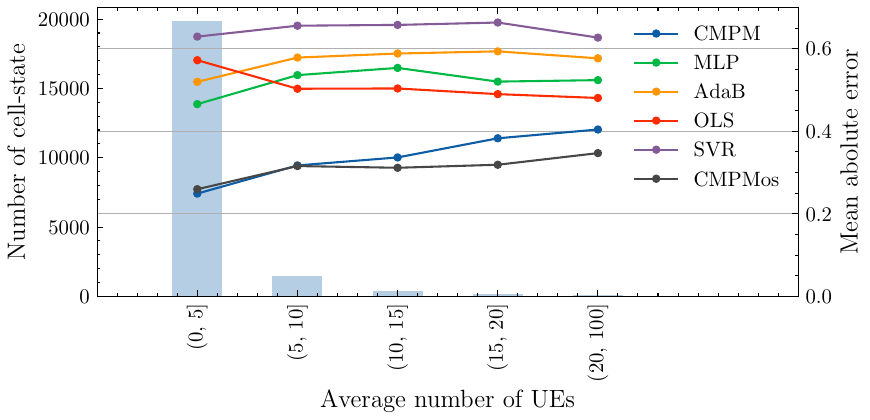}
\centering
\caption{Medians of MAE for different models (lines) vs average number of UEs. Bars denote number of cell-states with given number of cells range. Majority of cases falls into the lowest bin.}
\label{fig:n_cellstates_per_avgUEs_and_MAE}
\end{figure}

An interesting observation is made when the results are binned by the average number of active UEs for a given cell-state. Fig. \ref{fig:n_cellstates_per_avgUEs_and_MAE} presents the median values of mean absolute errors for each bin. For all cell state models (OLS, MLP, SVR and AdaBoost), the median of mean absolute errors remains fairly constant. For CMPM model, MAE values are increasing with the increase of average number of UEs. The model performance is simply better in regions, where there was more data available for training (see barplots in Fig. \ref{fig:n_cellstates_per_avgUEs_and_MAE}, which depicts the number of cell-states for each region of average number of UEs). There was only limited number of cells in the dataset working in high-load regime (average number of active UEs greater than 20), thus the model was not able to learn as effectively, as for low-loaded scenarios. This is important observation, marking the need of thorough design of sampling and training strategies for CMPM models. To prove that, we have trained additional CMPM model, denoted CMPMos in Fig. \ref{fig:n_cellstates_per_avgUEs_and_MAE}. During training, the datapoints for which number of UEs is higher than 10 has been oversampled. This simple strategy allowed to reduce MAE for groups with average number of UEs greater than 10. 

\begin{figure}
\includegraphics[width=3.4in]{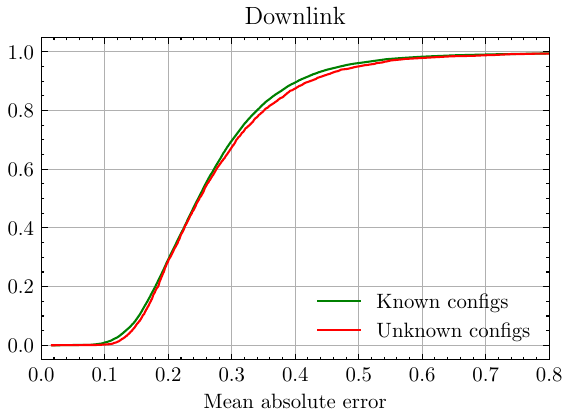}
\centering
\caption{CDFs of MAE for known and unknown configurations}
\label{fig:DL_CDF_MAE_NN_vs_ols_known_unknown_configs}
\end{figure}

\section{Conclusion}
This paper emphasized the importance of performance prediction based on configuration and environmental variables. We have outlined the processing pipeline needed to transform data from OSS into a dataset for training and evaluation of CMPM model. Even though we did not focus on searching for the best neural network architecture, we managed to show that the configuration-performance model can outperform models tailored for fixed configuration, which implies that the model was able to learn from the vast amount of data passed during training and learned to reflect configuration-environment-performance relation in a better way, than the dedicated models. The model can predict known and unknown configurations with similar quality (measured with MAE metric - Fig. \ref{fig:DL_CDF_MAE_NN_vs_ols_known_unknown_configs}). We also identified several open problems that should be addressed in future work. CMPM model is created for a single KPI and is fed with CM data, which can impact this KPI. Alternative approaches may be considered, e.g., creating a hierarchical model in the first stage with a broad range of configuration parameters, later used to build dedicated KPI models. This way, all KPIs will be modeled with unified input, which streamlines the addition of new models for new KPIs. 

The sampling strategy could be clarified to improve results for highly loaded cells, which make up only a tiny fraction of the learning set. The prediction error for this minority of cases is deteriorated compared to low-loaded cells.
    

\section*{Acknowledgment}

This research was funded in part by Nokia under EU grant no. POIR.01.01.01-00-1329/20 and in part by the Wroclaw University of Science and Technology under Implementation PhD programme and was supported by the statutory fund of the Department of Systems and Computer Networks, Wrocław University of Science and Technology.



%
\bibliographystyle{IEEEtran}

\bibliography{bibliography}


\end{document}